# Analysis of carbon content in direct-write plasmonic Au structures by nanomechanical scanning absorption microscopy

Miao-Hsuan Chien,[#,§] Mostafa M. Shawrav,[#,§,**] Kurt Hingerl,[‡] Philipp Taus,[†] Markus Schinnerl,[†] Heinz D. Wanzenboeck,[†] and Silvan Schmid [#,*]

[#] *Institute of Sensor and Actuator Systems, TU Wien, Gusshausstrasse 27-29, 1040 Vienna, Austria.*

[†] *Institute of Solid State Electronics, TU Wien, Gusshausstrasse 27-29, 1040 Vienna, Austria.*

[‡] *Center for Surface- and Nanoanalytics, Johannes Kepler University Linz, Altenbergerstr 69, 4040 Linz, Austria*

## Abstract

The determination of the chemical content is crucial for the quality control in high-precision device fabrication and advanced process development. For reliable chemical composition characterization, certain interaction volume of the target material is necessary for conventional techniques such as energy-dispersive X-ray spectroscopy (EDX) and electron energy loss spectroscopy (EELS). This remains however a challenge for nanostructures. We hereby propose an alternative technique for measuring chemical composition of nanostructures with limited volume. By measuring the differences in the optical absorption of the nanostructure due to the differences in the chemical composition with the resonance frequency detuning of a nanomechanical resonator as well as the assistance of the analytical optical modelling, we demonstrate the possibility of characterizing the carbon content in the direct-write focused electron beam induced deposition (FEBID) gold nanostructures.

## Introduction

Localized surface plasmon resonance (LSPR) is capable of creating gigantic field enhancements focused within subwavelength region, and has thus enabled a wide variety of applications in the recent decays, such as single-molecule Raman spectroscopy,[1,2] biomolecule detection,[3–5] high-efficiency solar cells,[6] as well as light-sensitive devices[7]. With LSPR, the plasmonic resonance frequency can be tailored by the geometry of the nanostructures. To fabricate well-defined plasmonic nanostructures, various nanofabrication techniques including interference lithography and electron beam lithography have been used.[8–10] However, these photoresist-based conventional techniques rely on a photoresist layer of homogeneous thickness and can only be applied on planar surfaces and depend on a complex sequence of multiple process steps. In order to fabricate complex plasmonic structures on arbitrary samples, an alternative flexible direct-write process for plasmonic materials is required. Focused electron beam induced deposition (FEBID) is such a direct-write approach that allows for a mask-free and resist-free deposition of plasmonic nanostructures.[11–17] In FEBID volatile precursor molecules of the desired material are injected into a scanning electron microscope through a gas injection system (GIS) and are locally deposited by the impinging electron beam.

---

[§] These authors contributed equally.
[**] Corresponding author for FEBID part: shawrav@gmail.com
[*] Corresponding author: silvan.schmid@tuwien.ac.at

FEBID direct write approaches for fabricating gold nanostructures have been presented recently.[18–20] It has been demonstrated, that complex 3D gold nanostructures deposited by FEBID display plasmonic properties.[20] Beside numerous benefits including direct patterning and 3D nanoprinting, the low purity of the noble metal structures deposited by FEBID is a major drawback of this technique. Generally, about 25-30 atomic % (at. %) gold is present in typical as-deposited FEBID gold structures. The rest of the chemical composition is mainly contributed by the residual carbon content from the deposition precursor. Therefore, various purification methods, such as electron beam exposure,[21] annealing,[22] laser assisted FEBID,[23] substrate heating,[24,25] post-deposition exposure to water,[26,27] and oxygen plasma[28,29] have been explored to increase the metal content of the structure recently. By optimizations of the FEBID process with an additional oxidizing gas, it has become possible to directly deposit chemically pure gold,[30] which is a prerequisite for obtaining a strong plasmonic resonance. In this work we report on a simple and effective post-deposition purification approach in which water acts as the oxidative gas species, purifying already deposited gold nanostructures. The bulk chemical composition of the structure is obtained using scanning electron microscopy (SEM) equipped with energy dispersive spectroscopy (EDX) and the morphology of the structure is measured using atomic force microscopy (AFM).

However, the conventional chemical composition analysis methods such as aforementioned EDX require a certain interaction volume of the sample material with the probing electron or radiation beam. For a direct-write plasmonic nanostructure, the volume is limited to nanoscale. As a result, in addition to the standard analysis methods, we introduce a new technique based on nanomechanical scanning absorption microscopy (NSAM),[31–33] which allows for the direct measurement of the chemical composition of individual plasmonic nanostructures. The NSAM method is based on the frequency detuning of a nanomechanical resonator due to the photothermal heating by plasmonic nanostructures. The optical absorption cross section of the nanostructures can be extracted from the frequency shift of the nanomechanical resonator precisely. Similar characterizations of the optical absorption of nanoparticles have also enabled the displacement detection down to picometers resolution.[34] In this study, the change in resonance frequency is directly correlated to the level of impurity within the single plasmonic nanostructures by the theoretical analysis based on effective medium approximation analysis and Mie theory. Hence, NSAM represents a viable alternative to electron energy loss spectroscopy (EELS), Auger electron spectroscopy and high-angle annular dark-field imaging. Finally, we use

NSAM to map and analyze purified FEBID Au bowtie antenna with various gap sizes to demonstrate the strong LSPR in the purified nanostructures.

## Results & Discussions

### I. Optimization of FEBID purification process

In order to investigate purification strategies and optimize the parameters, first Au nanostructure were deposited using a conventional FEBID procedure as shown in Figure 1a. The Au precursor is injected into the vacuum chamber (base pressure < 2 x $10^{-6}$ mbar) of a scanning electron microscope (SEM) via a gas injection system (GIS). The precursor is decomposed by secondary effects caused by electrons from the focused e-beam impinging on the surface. Various post-deposition purification strategies have been explored to purify gold nanostructures. Among them, the simplest way to purify gold nanostructures is annealing by extended irradiation with the focused electron beam. In addition, it has been shown that transferring the deposited structures to an environmental SEM and introducing water as an oxidative gas improves the purification significantly. However, this requires additional, specialized equipment. Here, we test the water-based post-deposition electron beam purification strategy in-situ without changing to an environmental SEM, but perform purification in the conventional SEM, where also the deposition was conducted.

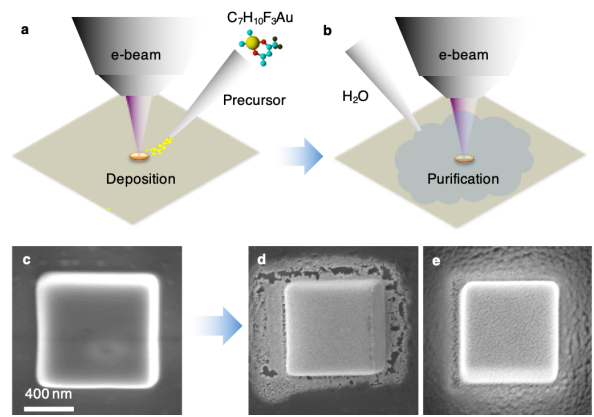

*Figure 1:* Schematics of (a) FEBID and (b) post-purification. SEM image of FEBID Au nanostructure (c) in pristine state after deposition, (d) after purification with continuous focused electron beam scanning in vacuum and (e) in $H_2O$ environment, respectively.

First, a series of Au rectangles as shown in Figure 1b. was deposited by FEBID. After the deposition, the structure was exposed to extensive electron beam irradiation (1 nA) at a pressure ~2.8 ×$10^{-6}$ mbar for 32 minutes. For the 1.2 x 1.2 μm² area of the Au-deposit this corresponds to 1,33 μC/μm². The exposure of the e-beam results in visible morphological changes on the surface. Second, in order to optimize an electron-beam activated purification with reactive species, deionized water was used as an oxidative additive. Though the reaction of carbon and water vapor to carbon oxide would happen spontaneously based on the Ellingham diagram, the electron-beam provides sufficient stimulus for overcoming the activation energy barrier of the reaction between carbon and $H_2O$. In this case, a freshly deposited structure was purified with the electron beam in the presence of water for 32 minutes at a pressure of $\sim 2.0 \times 10^{-4}$ mbar. After the purification, the surface looks smoother compared to e-beam curing alone.

While the morphological differences are apparent in Figure 1d and 1e, it is interesting to note that the EDX spectra differ even more (see Supplementary Figure S1); The EDX spectrum of the pristine deposit shows a significant count at the characteristic energy for carbon. The e-beam only purified structure exhibits a slightly lower carbon peak compared to a pristine deposit, which can result from the annealing effect and enhanced diffusion of carbon in gold. When water is added, the EDX shows that significant amount of carbon is removed due to the oxidative species formed by the electron beam. The chemical composition of the structures is shown in Table 1.

*Table 1:* Chemical composition obtained via EDX.

| Treatment | C (at. %) | Au (at. %) |
|---|---|---|
| As-deposited | 44 | 43 |
| E-beam | 45 | 49 |
| E-beam + $H_2O$ | 21 | 76 |

A different carbon content of the gold structures should also be reflected by a different plasmonic response of the structures. This was investigated by nanomechanical scanning absorption microscopy (NSAM) using a laser Doppler-vibrometer.

## II. Characterization of carbon content with NSAM and the effective medium modeling of dielectric function

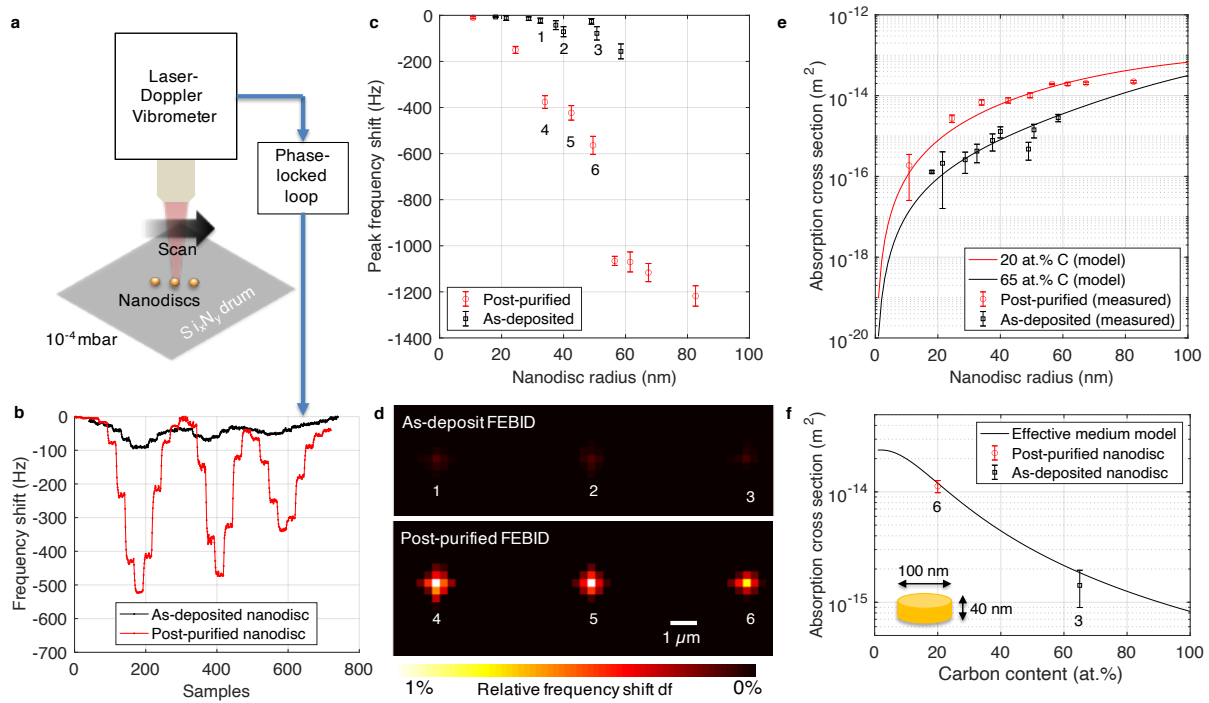

***Figure 2:*** *(a) Schematic of the nanomechanical scanning absorption microscopy (NSAM) measurement setup. (b) Typical raw frequency shift signals of NSAM measurements from the phase-locked loop. (c) The peak frequency shift of nanodiscs extracted from one-dimensional Gaussian fit, and (d) the corresponding NSAM mapping of as-deposited and post-purified FEBID nanodiscs from the labelled data points in (c), respectively. (e) The extracted absorption cross section from (c), and fitted by the absorption cross section calculated with Mie theory and dielectric function obtained from Maxwell-Garnett effective medium expression. (f) The absorption cross section of 100 nm gold nanodiscs under different content of carbon calculated from Mie theory and Maxwell-Garnett effective medium expression, with the measured absorption cross section of as-deposited (# 3) and post-purified (# 6) nanodiscs plotted as comparisons.*

In order to characterize the plasmonic properties of post-purified FEBID gold structures using NSAM, Au nanodiscs with different sizes were fabricated on multiple nanomechanical drums. The silicon nitride nanomechanical drums were prepared by the bulk micromaching lithography process. 50 nm silicon nitride thin film were first deposited with low-pressure chemical vapor deposition (LPCVD) process on the silicon substrate with ~150 MPa tensile stress, and subsequently released by a KOH etching process of silicon. Gold plasmonic nanostructures were then deposited on the drum with FEBID. Part of the Au anodiscs were purified using the previously introduced water assisted post deposition purification method. Both post-purified and as-deposited nanodiscs were analyzed with NSAM under high vacuum condition below a pressure of $10^{-4}$ mbar, as shown in Figure 2a. The mechanical resonance frequency of the silicon nitride drum was monitored with a laser-Doppler vibrometer (LDV). When the nanodiscs are scanned, the absorption of the laser by the FEBID nanodiscs results in a local heating of the drum, causing a thermoelastic detuning of the drum mechanical resonance frequency. This detuning of the resonance frequency is tracked by a phase-locked loop while scanning the drum, as shown in Figure 2a. The extend of the absorption of the laser depends on the absorption and plasmonic effects of the material, which – among other factors such as size and geometry - also depends on the purity of the Au nanodiscs. A test measurement signal with three as-deposited and three post-purified nanodiscs were plotted as in Figure 2b. In Figure 2b, the post-purified nanodiscs induce an almost 10-fold stronger frequency shifts than the as-deposited nanodiscs. This indicates a much higher optical absorption for purified nanostructures. The near-Gaussian frequency shift profile from the nanodiscs in Figure 2b results from the convolution of Gaussian laser beam and the absorption of the nanodisc, represented by a point absorber, as discussed in the previous studies.[35,36]

To investigate the effect of the carbon content on the optical absorption of the FEBID nanostructures systematically, the peak frequency shifts of nanodiscs with different sizes were extracted with a one-dimensional Gaussian fit, as plotted in Figure 2c. The root-mean-square fitting error indicated by the error bar. The frequency shift mapping of selected nanodiscs, labelled as #1 to #6 in Figure 2c, is plotted in Figure 2d. Nanodiscs #1 to #3 have similar radius with nanodiscs #4 to #6, and the only difference should be the purification process. The radius of the nanodiscs was measured by SEM, and the thickness was measured by AFM to be around 40 nm for such deposition parameter. To avoid the plasmon coupling between the nanodiscs and the grating effect, all the nanodiscs were deposited with at least 5 μm spacing. In general, from both Figure 2c and 2d, for nanodiscs with different radius, the post-purified nanodiscs demonstrated higher frequency shift with higher optical absorption, implying an enhanced LSPR for post-purified nanodiscs.

To quantify this effect of the carbon content on LSPR, the optical absorption cross sections $\sigma_{abs}$ of each nanodisc can be extracted respectively from drum frequency detuning $\Delta f$, namely

$$\sigma_{abs} = \frac{\frac{\Delta f}{f_0}}{RI} \quad (1)$$

with the previously measured responsivity $R$ of the silicon nitride drum of 850 W$^{-1}$, peak irradiance $I$ of 250 μW/μm$^2$, and $f_0$ is the reference resonance frequency of the silicon nitride drum to be around 52 kHz, measured by the LDV, as shown in Figure S2. The absorption cross sections were extracted from these vibrometer measurements and are summarized in Figure 2e. A goal of this study was to establish a theoretical relation between carbon content and absorption cross section and fit this theoretical model with the measured absorption cross section extracted from the frequency shift. In this way we can compare carbon content in the as-deposited and post-purified nanodiscs, and can extract the level of carbon content.

First, we consider the theoretical absorption cross section $\sigma_{abs}$ of individual gold nanosphere, which can be calculated based on the Mie theory absorption model [37]

$$\sigma_{abs} = \frac{2\pi}{\lambda} Im\{\alpha\} \quad (2)$$

with wavelength $\lambda$ and polarizability $\alpha$. Since the dipolar resonance behaviors are similar between nanospheres and nanodiscs along the axial polarization, this solution can also be used for our following analysis. For a homogeneous nanosphere of pure gold, the polarizability $\alpha$ for a dipolar mode is dependent on dielectric function of the gold material $\varepsilon$ and its surrounding medium $\varepsilon_m$, such that[38]

$$\alpha = 3V\varepsilon_m \frac{\varepsilon(\lambda) - \varepsilon_m}{\varepsilon(\lambda) + 2\varepsilon_m} \quad (3)$$

in which V is the volume of the nanosphere, $\lambda$ is the optical wavelength, $\varepsilon_m = 1$ for the case of vacuum. From equation (3), it can be clearly seen that the polarizability $\alpha$ and thus the theoretical absorption cross section $\sigma_{abs}$ of a nanosphere is dependent on the dielectric function of the gold material $\varepsilon(\lambda)$, For our FEBID nanodiscs the dielectric function material $\varepsilon(\lambda)$ is strongly dependent on the carbon content level of the gold nanodiscs.

Secondly, we need to consider the effect of carbon content on the effective dielectric function of the deposited gold nanodiscs. Due to the effect of interband transition in gold starting around the wavelength below 700 nm, modeling the dielectric function with Drude model by varying the electrical conductivity and damping rate would result in large discrepancy for the 633 nm wavelength used in present study.[39] To overcome this, also to provide a more general model independent of material and excitation wavelength, we apply the effective medium theory,[40] which obtain the effective dielectric function of the FEBID gold based on the dielectric function of pure gold,[41] pure graphite[42] and the level of impurity. By micro-Raman spectroscopy the carbon matrix in FEBID deposits was shown to be amorphous or graphite-like so that using the dielectric function pure graphite[42] was the closest match to model the carbon contents. For FEBID-Pt-deposition Poratti et al also describe the material composition as granular metal embedded in a carbonaceous matrix., the effective dielectric function $\varepsilon_{eff}$ of the FEBID gold with carbon content can be obtained by the Maxwell-Garnett effective-medium expression as in[40]

$$\varepsilon_{eff} = \varepsilon_{Au} \frac{2(1-p_c)\varepsilon_{Au}+(1+2p_c)\varepsilon_C}{(2+p_c)\varepsilon_{Au}+(1-p_c)\varepsilon_C}, \quad (4)$$

in which $\varepsilon_{Au}$ and $\varepsilon_C$ are the dielectric function of pure gold and pure graphite, respectively, and $p_c$ is the volume fraction of the carbon content. For the consistency with the EDX data, which is given in atomic fraction, the atomic fraction $a_c$ will be used in the following discussions. As indicated by AFM thickness measurements with our experiments no significant change in the volume with different carbon content was observed. Hence, the volume fraction $p_c$ can be translated to mass fraction $m_c$ and subsequently atomic fraction $a_c$ by

$$m_c = \frac{p_c D_c}{p_c D_c + (1-p_c) D_{Au}}, \quad (5)$$

$$a_c = \frac{\frac{m_c}{12}}{\frac{m_c}{M_C} + \frac{(1-m_c)}{M_{Au}}}, \quad (6)$$

where $D_c$ and $D_{Au}$ are the density of carbon and gold to be 2.26 and 19.30 g/cm³, respectively. $M_c$ and $M_{Au}$ are the atomic mass of carbon and gold to be 12.01 and 196.96 g/mol, respectively. The effective dielectric function $\varepsilon_{eff}$ obtained by the Maxwell-Garnett effective-medium expression is then applied to the Mie theory model in Equation (3),[40] and the absorption cross section of different carbon content can be calculated, as shown in the model in Figure 2e. By fitting this model to different sizes of nanodiscs instead of just single one, the individual deviations can be corrected to give a more robust picture, and the model can also be better verified.

As shown in Figure 2e, the measured absorption cross sections of the post-purified FEBID nanodiscs fit with the Maxwell-Garnett model of ~20 at.% of carbon content, which is quite consistent with the chemical composition obtained by EDX shown in Table 1. However, the measured absorption cross sections of the as-deposited FEBID nanodiscs fit with the model of ~65 at.% of carbon content, which deviates from the EDX value of 44 at.%. This deviation can be the fact that the Maxwell-Garnett effective medium expression assumes a host material with higher chemical composition and an incorporated material with lower chemical composition, which in our case are gold and graphite, respectively. As briefly mentioned, the incorporated material would form nanoscopically heterogenous domains in the matrix material. However, when the content of the carbon approaches the content of the gold, the heterogenous micro-domain picture of the incorporation in the Maxwell-Garnett approximation starts to show some discrepancies. Instead, a more randomly mixed microstructure with no clear differences between host and incorporated materials proposed by Bruggeman would be more appropriate for modeling highly carbon-contaminated gold structures.[40,43,44] As the present study intended to quantify small carbon content in plasmonic Au nanostructures, the Maxwell-Garnett model is clearly more suitable to meet the demands of this study. Numerical simulations can always provide more accurate modelling results; however this is beyond the scope of this discussion. The emphasis of this work is on the proof of physical concept. Another interesting observation in Figure 2e is that the measured absorption cross section of larger post-purified nanodiscs seems to deviate from the 20 at.% carbon content model to higher carbon content. This can result from the reduced surface to volume ratio for the larger nanodisc, so that the purification process was less effective with the same parameters.

To remove the contribution of the nanostructure geometry on absorption, measured absorption cross sections of as-deposited and post-purified nanodiscs both with 100 nm in diameter are plotted in Figure 2f. The calculated absorption cross sections for 100 nm nanodiscs from the proposed model with different levels of carbon content are also plotted in Figure 2f for comparison. The measured values fit well with the model at ~65 at. % and ~20 at. % for the as-deposited and post-purified. As shown in Figure 2f, the increased content of carbon would reduce the absorption cross section, implying a reduction in LSPR. Gold, as a low-loss plasmonic material, has a small imaginary part of the dielectric function in the visible regime. The contamination with graphitic carbon increases the imaginary part of the dielectric function and introduces additional damping to the plasmonic resonance. The proposed quantification analysis based on plasmonic absorption proves the reduction of carbon content in gold nanodiscs after the purification process aside from the conventional EDX, and also further provides a novel technique for the material characterization of nanostructures with a small volume for which a conventional point-probing technique would be impossible.

### III. LSPR in purified Au bowtie nanostructures

In order to show the enhanced plasmonic resonance in the post-purified FEBID nanostructures, as well as to show the versatility and potential of our technique for probing the absorption cross section, bowtie nanostructures with different gap sizes were deposited and post-purified, as shown in Figure 3a to c. From the finite-difference time-domain (FDTD) simulations shown in Figure 3d to f, it can be seen that the electromagnetic field enhancement scales inversely proportional to the gap size, as expected from existing models.[45] The same trend can be observed readily from the corresponding NSAM scans presented in Figure 3g to i.

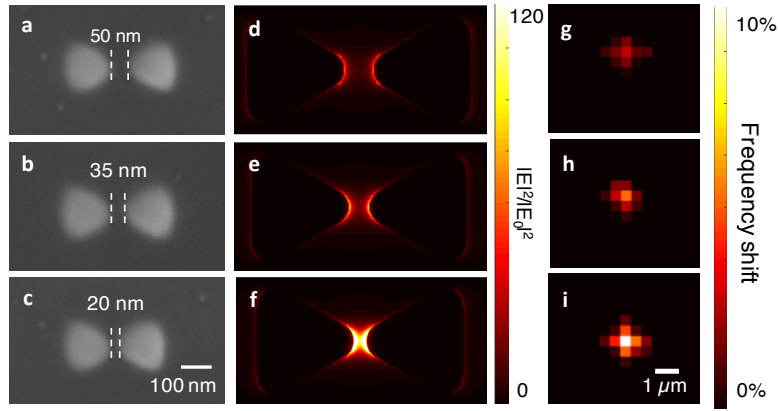

*Figure 3*: NSAM analysis of FEBID Au bowtie antennas. (a)-(c) SEM images of FEBID Au antennas with indicated gap sizes on top of silicon nitride. (d)-(f) Corresponding FDTD simulations of field intensities, and (g)-(i) corresponding NSAM scans of each antenna.

## Conclusions

This works shows that FEBID as a resistless direct-write method is capable of producing complex nanostructures even on fragile substrates with a high precision of tens of nanometers. We showed that post-deposition purification with $H_2O$ can significantly reduce the carbon content of the FEBID nanostructures and hence significantly enhances their plasmonic response. FEBID Au nanostructures of high purity opens the door to fast plasmonic prototyping of complex structures on non-standard surfaces. To investigate the carbon residuals in the as-deposited and post-purified nanostructures, we introduced nanomechanical absorption scanning microscopy (NSAM) as a new tool to quantitatively characterize chemical composition of single plasmonic nanostructures, more specifically the carbon content. NSAM combined with Mie theory and effective medium modeling can measure the chemical composition of individual Au nanostructures and allows us to analyze the FEBID purification process. Due to its high single particle sensitivity, NSAM has the potential to be a unique analysis tool especially for nanometer-sized samples and has potential applications in many fields and plasmonics in particular.

METHODS

All FEBID experiments were performed inside a Zeiss Leo 1530VP Scanning Electron Microscope which has a home-built multi-nozzle gas injection system (GIS). The base pressure of the system was ~ $2 \times 10^{-6}$ mbar. To deposit gold, commercially available dimethyl gold (III) trifluoroacetylacetonate precursor was used. The nozzles are cylindrical in shape with an inner diameter of 400 μm. The deposition area was approximately 1 mm away from the center of the nozzle outlet. Generally, during Au deposition, the gold precursor reservoir was heated up to ~50°C. in order to yield a working pressure of ~ 2 × 10$^{-5}$ mbar. In order to obtain reproducible results, the water was first outgassed in an ultrasonic bath. During the curing process the precursor reservoir was surrounded by a larger containment of water at room temperature, to compensate for evaporation cooling inside the reservoir caused by the exposure to vacuum.

EDX: EDX analysis was performed using Zeiss Neon 40Esb cross-beam microscope with an Oxford Instruments EDS 7427 detector. The obtained spectra were normalized to the gold peak, and the zero peak was removed.

Nanomechanical Scanning Absorption Microscopy: A Polytec laser Doppler vibrometer (MSA-500) was used to characterize the samples. The laser wavelength used in present study is 633 nm, the laser featured a spot size of ~1.5 μm. For this, a silicon nitride drum, which functions as a nanomechanical resonator with low background absorption, was used as sample.

Fabrication of silicon nitride drum: The samples are fabricated with a bulk micromaching process. A silicon wafer with a thickness of ~520μm is coated with 50 nm silicon-rich silicon nitride (SiN) with low pressure chemical vapor deposition (LPCVD). The release of the drum is done by a wet-etching process of the silicon in KOH for ~8 hours. The prestress of the released drum is approximately 150 MPa, which is extracted from the measured mechanical resonance frequency of the drum.

ACKNOWLEDGMENTS

Authors are grateful for the assistance of Sophia Ewert on device fabrication and Center for Micro and nanostructures (ZMNS) for the cleanroom support. This work has received funding from the European Research Council under the European Union's Horizon 2020 research and innovation program (Grant Agreement – 716087 – PLASMECS).